\documentclass[twocolumn]{aastex631}

\newcommand{\bu}{\mathbf{u}}

\newcommand{\rev}[1]{{\bf #1}}
\newcommand{\revout}{\bgroup\markoverwith{\textcolor{blue}{\rule[0.5ex]{2pt}{1pt}}}\ULon}

\begin{document}

\title{Constraints on the long-term existence of dilute cores in giant planets}
\author{A. Tulekeyev}
\affiliation{Department of Applied Mathematics, Baskin School of Engineering, University of California, Santa Cruz\\
1156 High St, Santa Cruz, CA 95064 USA}

\author{P.Garaud}
\affiliation{Department of Applied Mathematics, Baskin School of Engineering, University of California, Santa Cruz\\
1156 High St, Santa Cruz, CA 95064 USA}

\author{B. Idini}
\affiliation{Department of Astronomy and Astrophysics, University of California, Santa Cruz\\
1156 High St, Santa Cruz, CA 95064 USA}

\author{J. J. Fortney}
\affiliation{Department of Astronomy and Astrophysics, University of California, Santa Cruz\\
1156 High St, Santa Cruz, CA 95064 USA}

\begin{abstract}
  Ring seismology has recently revealed the presence of internal gravity waves inside Saturn that extend up to 60\% of Saturn's radius starting from the center, in what is recognized today as Saturn's stably-stratified dilute core. Similarly, gravity measurements on Jupiter suggest the existence of a dilute core of still poorly constrained radial extent.  These cores are likely in a double-diffusive regime, which prompt the question of their long-term stability. Indeed, previous DNS (Direct Numerical Simulations) studies in triply-periodic domains have shown that, in some regimes, double-diffusive convection tends to spontaneously form shallow convective layers, which coarsen until the region becomes fully convective. In this letter, we study the conditions for layering in double-diffusive convection using different boundary conditions, in which temperature and composition fluxes are fixed at the domain boundaries. We run a suite of DNS varying microscopic diffusivities of the fluid and the strength of the initial stratification.  We find that convective layers still form as a result of the previously discovered $\gamma$-instability which takes place whenever the local stratification drops below a critical threshold that only depends on the fluid diffusivities. We also find that the layers grow once formed, eventually occupying the entire domain. Our work thus recovers the results of previous studies, despite the new boundary conditions, suggesting that this behavior is universal.
  The existence of Saturn's stably-stratified core, today, therefore suggests that this threshold has never been reached, which places a new constraint on scenarios for the planet's formation and evolution.  
\end{abstract}


\section{Introduction}\label{sec:intro}

Simple formation scenarios via the ``core accretion'' mechanism predict the existence of a compact mass of heavy elements concentrated near the center of Jupiter and Saturn, in what is known as the traditional view of gas giant planet interiors \citep{safranov1969evolution,perri1974hydrodynamic,mizuno1978instability,mizuno1980formation,pollack1996formation,guillot2005interiors}. Over the past several years, new high-precision constraints on the gravity field of Jupiter from the Juno Mission \citep{bolton2017juno,folkner2017jupiter,durante2020jupiter}, and the gravity field and seismic oscillation frequencies of Saturn from the Cassini Mission \citep{hedman2013kronoseismology,hedman2014more,hedman2018kronoseismology,french2019kronoseismology,iess2019measurement}, have revolutionized our  understanding  
of the interior structure of these planets, 
providing observational evidence that puts this traditional view into question. A wealth of evidence suggests that these planets have a dilute or fuzzy core that extends up to at least ${\sim}50\%$ 
of the planetary radius in the case of Saturn \citep{fuller2014saturn,MankovichFuller2021}, and up to 30\% to 70\% of the radius in the case of Jupiter \citep{wahl2017comparing,debras2019new,miguel2022inhomogeneous,idini2022gravitational,militzer2022juno,howard2023jupiter}.  
These cores are composed of a mixture of heavy elements and H/He fluid, with an abundance of heavy elements that increases towards the planetary center, thus creating a strongly stabilizing composition gradient.

The case for Saturn is especially compelling, as Saturn's interior is constrained not just by the gravity field but also by ring seismology \citep{marley1993planetary,fuller2014saturn,mankovich2020saturn,MankovichFuller2021}.  \citet{MankovichFuller2021} 
have recently found strong evidence for a dilute core structure that is stably-stratified, meaning that it admits internal gravity wave normal modes, and is not freely convecting.  
 Similar seismic constraints are not available for Jupiter,
but we have 
instead more precise measurements of the gravity field thanks to the Juno mission. All interior models published to date that fit this gravity field, by different groups using several different high pressure equations of state for H/He mixtures, require a dilute core \citep{miguel2022inhomogeneous,idini2022gravitational,militzer2022juno,howard2023jupiter}. However, the true extent of Jupiter's  core, and whether it is statically stable today, remains unclear due to the relatively large uncertainties in our knowledge of the equation of state for H-He and how they complicate the interpretation of Jupiter's gravity field  \citep{howard2023jupiter}.

Interior models that aim to reconstruct the planet's interior structure from the gravity field or ring seismology are inherently static models.  Furthermore, such models typically have little power to constrain the thermal structure of the planetary interior. However, additional information and constraints can be obtained by taking into account the stability of the fluid to turbulent motions and the effect that mixing caused by these motions would have on the heavy element distribution. For instance, while static models may allow a dilute core to support a strong compositional gradient, that solution could be unphysical from a fluid dynamical perspective  depending on the strength of the temperature gradient. Indeed, should the latter be strong enough to drive some form of convection despite the stabilizing compositional gradient, the turbulent motions 
could rapidly homogenize the heavy elements throughout the core. In that respect, it is therefore crucial to establish the stability of dilute cores to convective mixing, either via overturning convection or double-diffusive convection.


In what follows, we therefore study the properties of a Saturn-like dilute core, with a destabilizing temperature gradient, but a strongly stabilizing heavy-element gradient (so that the core overall is \emph{stable} to the Ledoux criterion).
This type of stratification can nevertheless be unstable to oscillatory double-diffusive convection (ODDC hereafter) in some cases, as well as secondary convective instabilities via layering (see more on this below). ODDC, which is often also called semi-convection, is a form of weak convection triggered by diffusively destabilized gravity waves \citep{Walin1964,Kato1966,baines1969}. It is primarily governed by three nondimensional parameters: the Prandtl number $Pr$, the diffusivity ratio $\tau$, and the inverse density ratio $R^{-1}$ \citep{rosenblumal2011}, defined respectively as:
\begin{equation}
    Pr = \frac{\nu}{\kappa_{T}}, \tau = \frac{\kappa_{C}}{\kappa_{T}},  R^{-1} = \left|\frac{N^{2}_{C}}{N^{2}_{T}}\right|,
\end{equation}
where $\nu$ is the kinematic viscosity of the fluid, $\kappa_{C}$ and $\kappa_{T}$ are compositional and thermal diffusivities, respectively, and $N^{2}_{T}$ and $N^{2}_{C}$ are the Brunt-V\"ais\"al\"a frequencies due to the thermal and compositional stratification, respectively. While the Prandtl number and the diffusivity ratio characterize properties of the fluid, the inverse density ratio, $R^{-1}$, defines the strength of stratification, with more strongly stratified regions having a larger $R^{-1}$. The criterion for linear instability to ODDC is \citep{Walin1964,Kato1966}:
\begin{equation}
    1\le R^{-1}\le \frac{Pr +1}{Pr+\tau} \rev{\equiv} R_{C}^{-1}.
    \label{eq:oddc_crit}
\end{equation}
The limit $R^{-1} = 1 $ corresponds to the Ledoux criterion, and a region of fluid with $R^{-1}\leq 1 $ is unstable to overturning convection. A region with  $R^{-1}\geq R_{C}^{-1}$ on the other hand is linearly stable to ODDC. 

In general ODDC-induced  transport is weak, characterized by a total\footnote{By total here we mean the sum of the turbulent flux and the diffusive flux.} temperature flux $F_{T}$ that is at most ten times larger than the radiative flux, and a total compositional flux $F_{C}$ that is at most a hundred times larger than pure diffusion \citep[see figure 4 of][]{Mirouh2012}. However, direct numerical simulations (DNS) have shown that ODDC can also transition into layered convection under some circumstance \citep{rosenblumal2011}, and the fluxes in the layered state can be orders of magnitude larger than in the ODDC state \citep{Woodal13}. The layering transition is caused by the $\gamma$-instability, first discovered by \citet{radko2003mechanism} in context of fingering convection in the Earth's ocean, and applied to ODDC by  \citet{rosenblumal2011} and \citet{Mirouh2012}. The $\gamma$-instability can cause the growth of large-scale density anomalies  in double-diffusive convection (ODDC or fingering convection) because temperature and composition are not transported at the same rate in these small-scale instabilities, but both contribute to the total density.

To model this instability, we usually define the so-called `inverse flux ratio' $\gamma^{-1}$ as \footnote{This corresponds to $\gamma_{\rm tot}^{-1}$ in \citet{Mirouh2012} but we omit the subscript `tot' to simplify the notation.} :
\begin{equation}
    \gamma^{-1} = \frac{F_{C}}{F_{T}}.
\end{equation}
In homogeneous ODDC, $\gamma ^{-1}$ is only a function of $Pr$, $\tau$, and $R^{-1}$. 
\citet{rosenblumal2011} and \citet{Mirouh2012} showed analytically and numerically that the $\gamma$-instability only takes place when $\gamma^{-1}$ is a decreasing function of $R^{-1}$. For a given set of fluid parameters, the function $\gamma^{-1}(R^{-1})$ is always concave \citep[see for instance figure 5 of][]{Mirouh2012}, decreasing from $R^{-1}=1$ to a global minimum at $R_{L}^{-1}$, and then increasing again at larger $R^{-1}$. The position of this minimum depends on $Pr$ and $\tau$. This implies that convective layers are expected to form by the $\gamma$-instability whenever
 \begin{equation}\label{eq:r_l}
     1 \le R^{-1}< R_{L}^{-1}(Pr,\tau).
 \end{equation}
As shown by \citet{Mirouh2012} and \citet{Woodal13}, these convective layers are initially shallow but seem to always rapidly merge to occupy the entire thermally unstable region. The implication of these results, as discussed in \citet{Woodal13}, is that ODDC-unstable regions are not necessarily long-lived: if $R^{-1}$ drops below $R_{L}^{-1}$, the $\gamma$-instability rapidly triggers the formation of shallow layers that then gradually merge into a large-scale, chemically homogeneous convection zone. 

However, it is important to note that all of these previous studies were very idealized. In particular, they assumed triply periodic boundary conditions, in which the thermal and compositional fluxes in the domain can increase freely in response to layer mergers \citep{Woodal13}. By contrast, in a real planet these fluxes would be set by processes that take place on the much larger planetary lengthscale and on the evolutionary timescale. As a result, an important question is whether the shallow convective layer formation and gradual mergers happen as described above if the fluxes are externally fixed instead.  



In giant planets, for instance, the temperature flux at a given radius is due to the loss of thermal energy, initially generated from the planet's formation and subsequent additional gravitational contraction. The local temperature flux, up to the surface that radiates to space, therefore depends on the integrated local heat capacity and contraction rate, such that values nears the planet's center will necessarily differ from that at the surface.\citep[e.g.,][their Figure 17]{Nettelmann2015}. The compositional flux in the center arises by convective erosion of the compact core while the one at the surface is zero. These realistic boundary conditions could, in principle, be modelled in a full-planet simulation. However, this is not computationally possible owing to the high resolution required to capture the dynamics of ODDC, so we are forced to model a small region within the dilute core instead, of extent at most a few hundred meters in each direction \citep{MollGaraud2017}. The question of which boundary conditions to apply at the edge of the computational domain is much less clear. So, in this study,  we assume for simplicity that fluxes of temperature and composition are fixed in time, i.e. so-called `fixed-flux' boundary conditions, and are the same at both boundaries. While this is not necessarily more realistic than periodic boundary conditions, the comparison with previous works will establish which results are robust and which are model-dependent.

 In section \ref{sec:setup}, we present the model setup and the numerical method of solution. In section \ref{sec:results}, we present a suite of selected simulations that span a range of initial stratifications and analyze the results in the context of the $\gamma$-instability theory. Finally, in the section \ref{sec:discussion}, we discuss the implications of our results for planet formation and evolution models.

\section{Model setup} \label{sec:setup}
For the purpose of our study, we consider a Cartesian  domain of height $H$ under uniform gravity ${\bf g} = -g {\bf e}_{z}$. We assume that the domain is bounded by impermeable parallel plates at $z = -H/2$ and $z = H/2$, and is horizontally periodic in both $x$ and $y$ directions.
The equations are the usual Boussinesq equation for compressible gases \citep{SpiegelVeronis1960} which are valid as long as fluid velocities are much smaller than the local sound speed and $H$ is much smaller than the density, pressure, or temperature scale heights\footnote{Note that this implies that our domain only spans a small portion of the core.}. These are: 
\begin{eqnarray}
&& \nabla \cdot \bu = 0 , \label{eq:dim_divu}\\
&& \frac{\partial \bu}{\partial t} + \bu \cdot \nabla \bu  =  - \frac{1}{\rho_m} \nabla p + (\alpha_T T - \alpha_C C) g{\bf e}_z  +  \nu \nabla^2 \bu, \label{eq:dim_mom} \\
&&  \frac{\partial T}{\partial t} + \bu \cdot \nabla T - w \frac{dT_{\rm ad}}{dz} =  \kappa_T \nabla^2 T, \label{eq:dim_T} \\
&& \frac{\partial C}{\partial t} + \bu \cdot \nabla C =  \kappa_C \nabla^2 C, \label{eq:dim_C}
\end{eqnarray}
where $\bu = (u,v,w)$ is the velocity field, $p$ is the pressure perturbation away from hydrostatic equilibrium, $T$ is the temperature perturbation away from the mean temperature $T_{m}$ of the domain, and  $C$ is the composition perturbation away from the mean composition $C_{m}$. Note that the composition $C$ is simply the mass fraction of heavy elements (usually called $Z$), but we use $C$ here to avoid any confusion with the coordinate $z$. 
We assume that the fluid has a constant mean density $\rho_{m}$, kinematic viscosity $\nu$, thermal diffusivity $\kappa_{T}$, and compositional diffusivity $\kappa_{C}$. It also has constant coefficients of thermal expansion and compositional contraction, $\alpha_{T}$ and $\alpha_{C}$ respectively, defined as:
\begin{equation}
    \alpha_{T} = {\bf -} \left.\frac{1}{\rho_{m}} \frac{\partial \rho}{\partial T}\right|_{\rho_{m},C_{m}}, \quad 
    \alpha_{C} =\left. \frac{1}{\rho_{m}} \frac{\partial \rho}{\partial C}\right|_{\rho_{m},T_{m}}.
\end{equation}
Finally, $dT_{\rm ad}/dz$ is the adiabatic temperature gradient and is also assumed to be constant. 

Both plates are assumed to be no-slip. We impose fixed flux boundary conditions, namely, a positive temperature flux $F_{T}$, and a positive composition flux $F_{C}$ at each plate so that:
\begin{equation}
\label{eq:ffbc}
\frac{\partial T}{\partial z} = -\frac{F_T}{\kappa_{T}}
\mbox{ and } \frac{\partial C}{\partial z} = -\frac{F_{C}}{\kappa_C}
\end{equation}
at $z = \pm H/2$. In the absence of fluid flow, the steady-state solutions of (\ref{eq:dim_T}) and (\ref{eq:dim_C}) with boundary condition (\ref{eq:ffbc}) have constant temperature and composition gradients $dT_b/dz = -F_T/\kappa_T$ and $dC_b/dz = -F_C/\kappa_C$ throughout the domain. We then define the fluctuations away from that steady background, $\tilde{T}$ and $\tilde{C}$, such that :
\begin{equation}
T = z \frac{dT_{b}}{dz} +  \tilde{T}, \quad 
C = z \frac{dC_{b}}{dz} +  \tilde{C}. 
\end{equation}
The new boundary conditions are then:
\begin{equation}
 \frac{\partial \tilde{T}}{\partial z} = 0,  \mbox{ and } \frac{\partial \tilde{C}}{\partial z} = 0 
 \end{equation}
 at each plate, and the perturbation equations are
\begin{eqnarray}
&& \nabla \cdot \bu = 0, \nonumber\\
&& \frac{\partial \bu}{\partial t} + \bu \cdot \nabla \bu  =  - \frac{1}{\rho_m} \nabla \tilde{p} + (\alpha_T \tilde{T} - \alpha_C \tilde{C}) g{\bf e}_z  +  \nu \nabla^2 \bu, \nonumber\\
&&  \frac{\partial \tilde{T}}{\partial t} + \bu \cdot \nabla \tilde{T} + w   \frac{dT_b}{dz} - w \frac{dT_{\rm ad}}{dz}  =  \kappa_T \nabla^2 \tilde{T}, \nonumber \\
&& \frac{\partial \tilde{C}}{\partial t} + \bu \cdot \nabla \tilde{C} + w   \frac{dC_b}{dz}=  \kappa_C \nabla^2 \tilde{C}. 
\end{eqnarray} 

Next, we use the standard nondimensionalization for double-diffusive convection \citep{radko2013double}, where the unit length $[l]$, the unit time $[t]$, the unit velocity $[u]$, the unit temperature $[T]$ and the unit composition $[C]$ are given by
\begin{eqnarray}
&& [l] = d  = \left( \frac{\kappa_T \nu}{\alpha_T g | dT_b/dz -dT_{\rm ad}/dz| } \right)^{1/4},   \\ 
&& [t] = \frac{d^2}{\kappa_T}, \\
&&[u] = \frac{\kappa_T}{d}, \\
&&[T] = d | dT_b/dz - dT_{\rm ad}/dz |,   \\
&& [C] = \frac{\alpha_T}{\alpha_C} d | dT_b/dz -dT_{\rm ad}/dz |.
\end{eqnarray}
In these new units, the top boundary is located at the nondimensional height $z = H / 2d = \hat H /2$ , and the bottom boundary at $z = - H / 2d = -\hat H / 2$. The nondimensional boundary conditions become: 
\begin{equation}
    \hat u = \hat v = \hat w = 0, \quad \frac{\partial \hat{T}}{\partial z} =0 , \mbox{ and } \frac{\partial \hat{C}}{\partial z} = 0,
\end{equation}
at $z = \pm \hat H/2$, where hats are used to denote nondimensional dependent variables.

The nondimensional equations are:
\begin{eqnarray}
&& \nabla \cdot \hat \bu = 0, \\
&& \frac{\partial \hat \bu}{\partial t} + \hat \bu \cdot \nabla \hat \bu  =  - \nabla \hat{p} + Pr ( \hat{T} - \hat{C}) {\bf e}_z  +  Pr \nabla^2 \hat \bu, \\
&&  \frac{\partial \hat{T}}{\partial t} + \hat \bu \cdot \nabla \hat{T} - \hat w    =  \nabla^2 \hat{T}, \\
&& \frac{\partial \hat{C}}{\partial t} + \hat \bu \cdot \nabla \hat{C} - R_0^{-1} \hat w   = \tau \nabla^2 \hat{C}, 
\end{eqnarray} 
where the independent variables $(x,y,z,t)$ are now implicitly nondimensional as well, but the hats are omitted to avoid crowding the equations. The usual parameters for double-diffusive convection emerge, namely
\begin{equation}
Pr = \frac{\nu}{\kappa_T}, \quad \tau = \frac{\kappa_C}{\kappa_T}, \quad R_{0}^{-1} = \frac{\alpha_{C}  | dC_b/dz |}{\alpha_{T} | dT_b/dz - dT_{ad}/dz|}.
\end{equation}
The Prandtl number, $Pr$, is typically of order $10^{-2}$ or $10^{-1}$, while the diffusivity ratio $\tau$ is often an order of magnitude smaller than $Pr$ \citep{french2019kronoseismology}. Both are fluid properties. By contrast, $R_{0}^{-1}$ depends on the thermal and compositional stratification of the fluid, and for the simulations presented here, is the inverse density ratio at the start of the simulation. 

The initial conditions are:
\begin{equation}
    \hat{\bf u}= 0, \quad \hat{T}= \rev{0} 
    ,\quad \hat{C}= \rev{\epsilon}, 
\end{equation}
where $\epsilon$ is a small amplitude random noise.
To study this model numerically, we use 'Dedalus', an open-source Python library for spectral methods \citep{BurnsetalDedadlus}. The computational domain is Cartesian, and configured to use Fourier bases in both horizontal $x$ and $y$ directions, and a Chebyshev basis in the vertical direction $z$. The nondimensional domain size is $100 \times 100$ in the horizontal plane, and is of height $\hat H$ = 200 vertically.

\section{Simulations results} \label{sec:results}
We ran a number of simulations with two different values of the Prandtl number $Pr$ and diffusivity ratio $\tau$. 
In this section, we present a number of sample simulations at $Pr = \tau = 0.1$, that demonstrate typical behaviors and dynamics observed in all performed simulations listed in Table \ref{tab:simulations}. In each case, we show the evolution of the total nondimensional mean density profile, $\bar{\rho}(z,t)$, defined as :
\begin{equation}
    \bar{\rho}(z,t) = \frac{1}{2\Delta t L_{x} L_{y}}\int_{t-\Delta t}^{t+\Delta t} \int_{0}^{L_{x}} \int_{0}^{L_{y}} \hat{\rho}(x,y,z,t) dxdydt
\end{equation}
where $\hat \rho = - (\hat T - z) + (\hat C  - R_0^{-1} z)$ is the total nondimensional density, and $\Delta t \simeq 50$. The time integral is useful to filter out fast gravity waves dynamics. We also compute the local inverse density ratio, $R^{-1}(z,t)$, defined as:
\begin{equation}\label{eq:localR}
    R^{-1}(z,t) = \frac{\frac{\partial \bar{C}(z,t)}{\partial z}}{\frac{\partial \bar{T}(z,t)}{\partial z}},
\end{equation}
where $\bar{C}$ and $\bar{T}$ are defined similar to $\bar{\rho}$. 

We find that our simulations fall into 4 categories described below. 



\subsection{Highest  $R_0^{-1}$}

\begin{figure*}
\centerline{\includegraphics[width=0.9\textwidth]{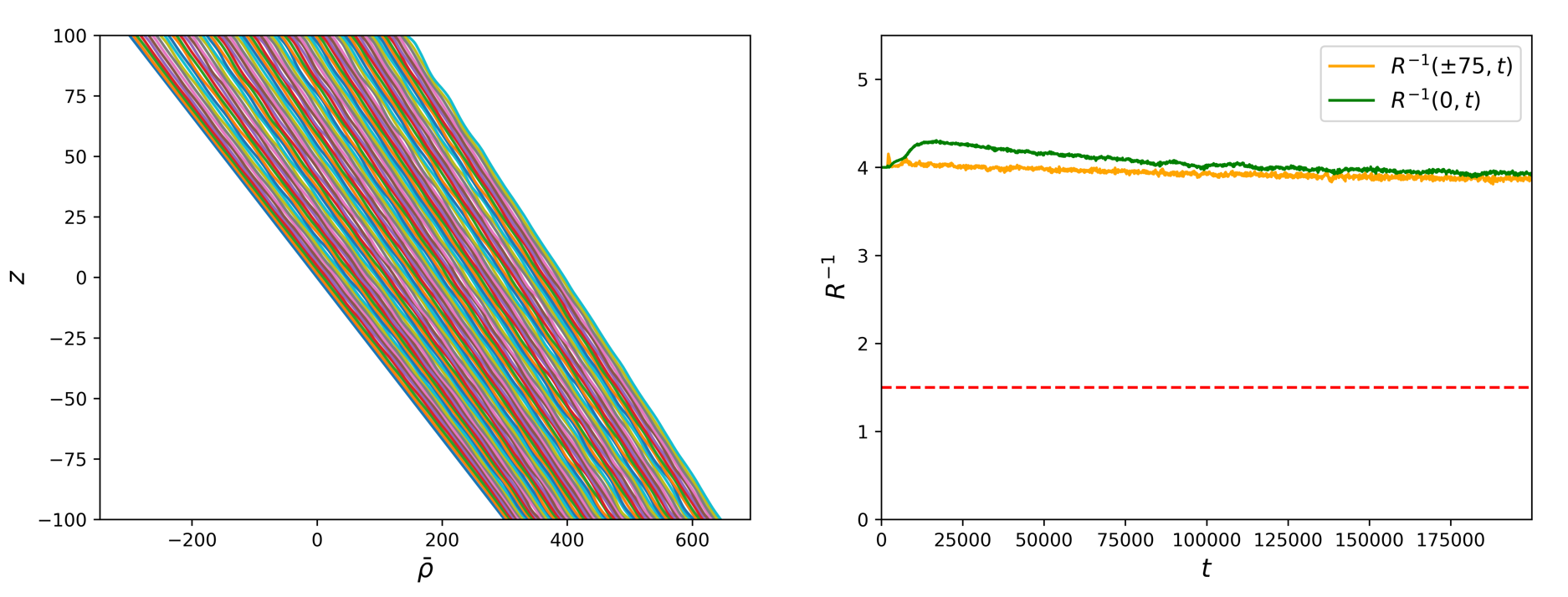}}
    \caption{Left: Time-series of nondimensional mean vertical density profiles (with $\bar \rho$ on the horizontal axis and the $z$-coordinate on the vertical axis) for $Pr = \tau = 0.1$ and $R^{-1}_{0} = 4.0$. Each line represents a mean density profile at a specific time, and each density profile is horizontally offset (shifted) from earlier time steps for ease of visualization. Lines are plotted every 2500 time units increments, in 10 different colors. Each blue line corresponds to $\Delta t = 25,000$ time unit increments. Right: Inverse density ratios (vertical axis) vs time (horizontal axis): $R^{-1}(0,t)$ (green), and $R^{-1}(\pm 75, t)$ (orange). The red horizontal dashed line shows $R_L^{-1} \simeq 1.5$, which is the layering threshold for $Pr = \tau = 0.1$.}
    \label{fig:case1}
    \end{figure*}

We begin with the simplest case which takes place at $R_{0}^{-1}$ close to, but still smaller than $R_C^{-1}$. In this case, the fluid becomes unstable to ODDC, and the system eventually reaches a statistically-stationary ODDC state. Figure \ref{fig:case1} (left) represents a time series of density profiles for the simulation with $Pr = \tau = 0.1$ and $R_{0}^{-1} = 4.0$ which is illustrative of this regime, i.e. a strong stratification, but still unstable to ODDC by the criterion given in equation (\ref{eq:oddc_crit}). The vertical axis corresponds to the $z$-coordinate. The simulation starts with a linear density profile at $t = 0$. Throughout the evolution, we observe weak perturbations from the linear background, but overall the shape and slope of the density profiles remain approximately the same. In Figure \ref{fig:case1} (right), we show the time evolution of $R^{-1}(0,t)$, namely the instantaneous inverse density ratio in the middle of the domain, as well as $R^{-1}( \pm75,t)$, the  inverse density ratio closer to the boundaries. We see that both curves start at the initial value $R^{-1}_{0}$, increase slightly at early times as a result of ODDC, and later converge asymptotically to the same constant value as the simulation reaches a statistically-stationary state.

\subsection{High $R_0^{-1}$}
\begin{figure*}
\centerline{\includegraphics[width=0.9\textwidth]{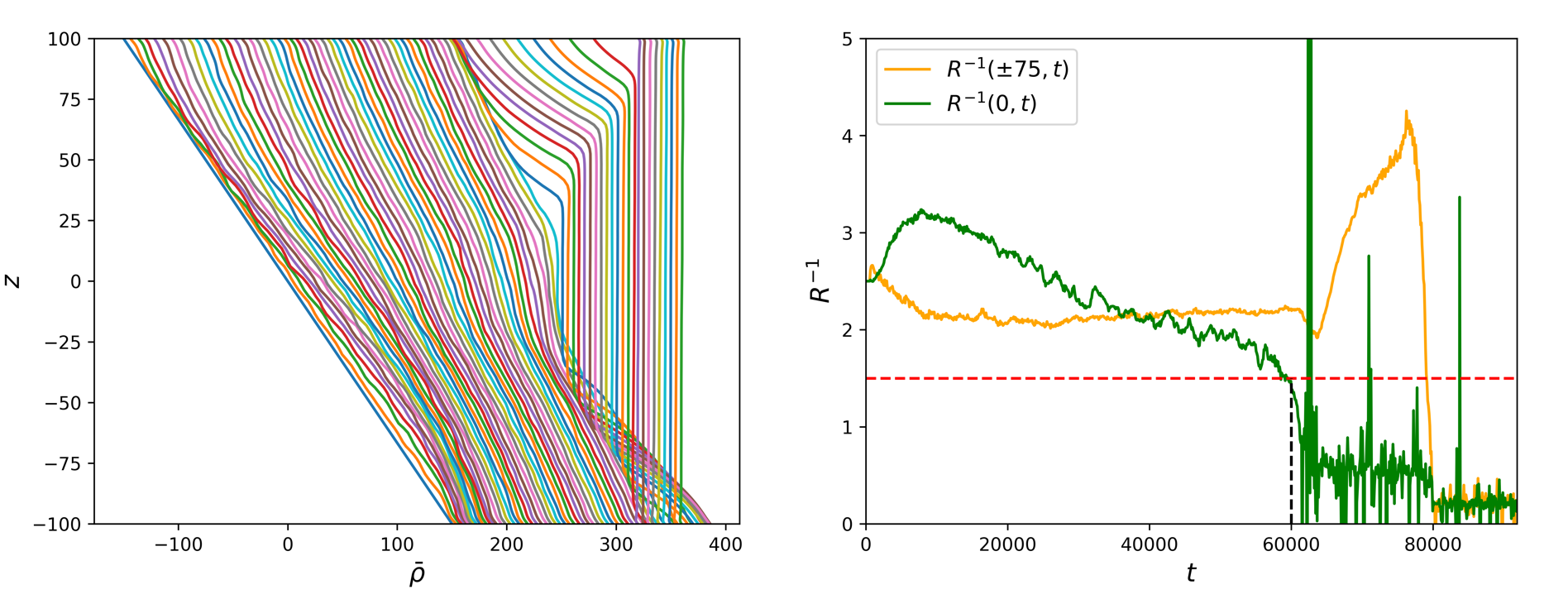}}
    \caption{As in Figure \ref{fig:case1}, for $Pr = \tau = 0.1$ and $R^{-1}_{0} = 2.5$. Left: Time-series of vertical density profiles. Each blue line corresponds to $\Delta t = 12,500$ time unit increments. Right: Inverse density ratios vs time: $R^{-1}(0,t)$(green), and $R^{-1}(\pm 75, t)$ (orange). The red horizontal dashed line shows $R_L^{-1} \simeq 1.5$, which is the layering threshold for $Pr = \tau = 0.1$, and the black dashed line corresponds to $t = 60,000$ when convection is triggered.}
    \label{fig:case2}
    \end{figure*}

Next we look at a slightly lower $R_{0}^{-1}$. Figure \ref{fig:case2} shows the same diagnostics as Figure \ref{fig:case1}, for parameters $Pr = \tau = 0.1$, $R_{0}^{-1} = 2.5$. We see that the domain also becomes unstable to ODDC at early times, and stays in that state for a very long time. In contrast with Figure \ref{fig:case1}, however, it does not reach a statistically-stationary ODDC state. We see that as the ODDC develops, $R^{-1}(0,t)$ decreases with time, which means that the background density stratification becomes weaker. Eventually around $t\approx60,000$, we see a clear convective layer appear in the middle of the domain, characterized by a constant density. 
To understand why a convective layer form, we note that $R^{-1}(0,t)$ crosses the $\gamma$-instability threshold $R^{-1}_{L}$, shown as the horizontal dashed red line in Figure \ref{fig:case2} \citep[which is approximately equal to 1.5 at these parameters, see][]{Mirouh2012} at $t\approx 60,000$. Shortly thereafter, the $\gamma$-instability causes an inversion in the mean density profile, which eventually overturns into convection. Later, this convective layer grows to fill the entire domain around $t\approx 75,000$, and the system is now in a statistically-stationary fully-convective state. The mechanism by which the convective layer grows will be explored in a future publication. 

\subsection{Intermediate $R_0^{-1}$}

\begin{figure*}
\centerline{\includegraphics[width=0.9\textwidth]{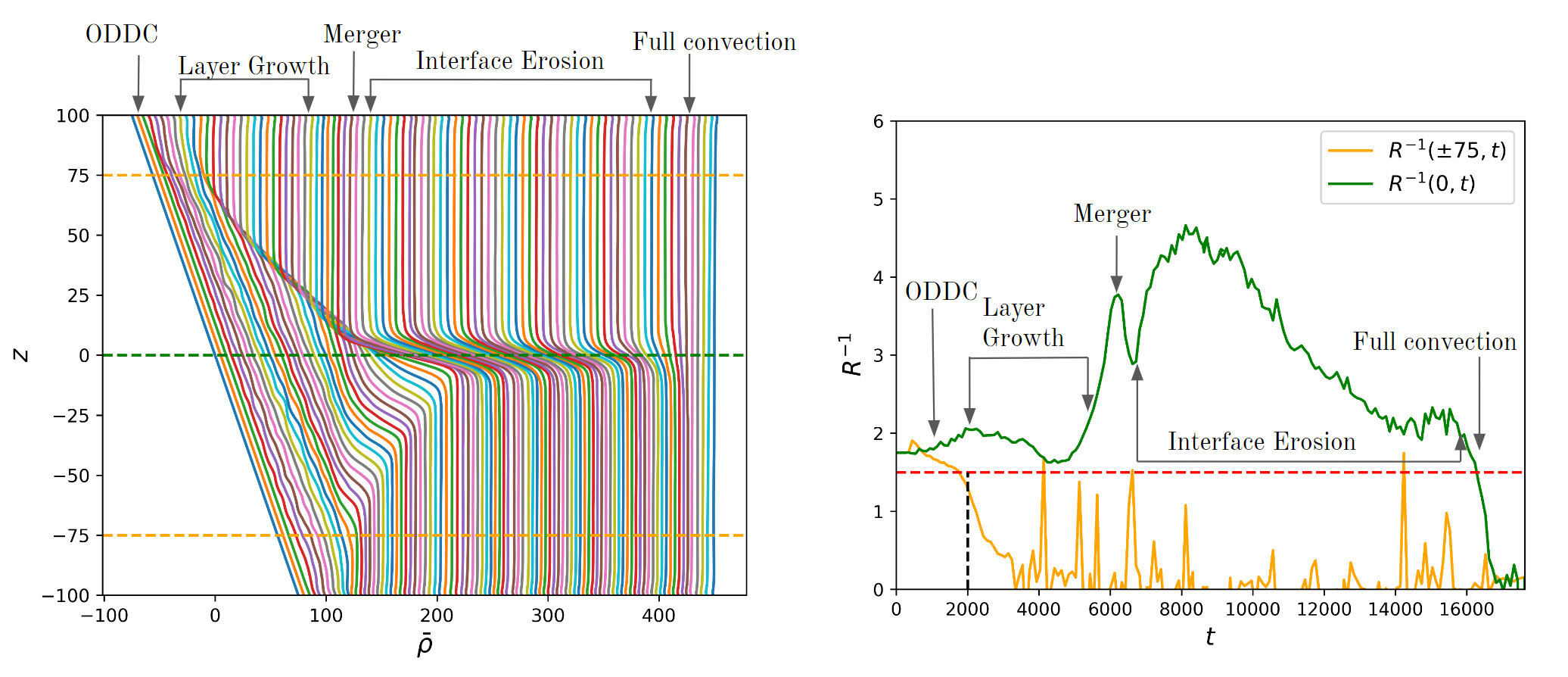}}
    \caption{As in Figure \ref{fig:case1}, for $Pr = \tau = 0.1$ and $R^{-1}_{0} = 1.75$. Left: Time-series of vertical density profiles. Each blue line corresponds to $\Delta t = 2000$ time unit increments. The various evolutionary stages are marked with arrows. Right: Inverse density ratios vs time: $R^{-1}(0,t)$(green), and $R^{-1}(\pm75,t)$ (orange). The red horizontal dashed line shows $R_L^{-1} \simeq 1.5$, which is the layering threshold for $Pr = \tau = 0.1$, and the black dashed line corresponds to $t=2000$ when the convection is triggered.}
    \label{fig:case3}
    \end{figure*}

As we continue to weaken the initial stratification (i.e. lower $R_0^{-1}$), we find that the long-term evolution of the system is very different, and notably, that convective layers form from the boundaries of the domain, rather than from the middle. This is illustrated in Figure \ref{fig:case3} where we present results for a simulation with $Pr = \tau = 0.1$ and $R_{0}^{-1} = 1.75$. As usual, the simulation starts with a linear density profile at $t = 0$ and rapidly becomes unstable to ODDC. This causes weak deviations from the linear profile observable around $t \approx 1000$. However, in this case, we can see convective layers emerge near the boundaries of the domain around $t \approx 2000$. Inspection of $R^{-1}(\pm 75,t)$ in Figure \ref{fig:case3} (right) reveals that this time $R^{-1}(z,t)$ drops below $R_{L}^{-1}$ near the domain boundaries first. Meanwhile, $R^{-1}(0,t)$ remains above $R^{-1}_{L}$, so the intermediate region remains in the state of ODDC, at least temporarily. These boundary convective layers grow in time between $t\approx 2000$ and $t\approx 6000$ through an erosion process that appears to be consistent (at least qualitatively) with that observed by \citet{Fuentes2022} in related simulations. After some time, the intermediate ODDC region completely disappears, and is replaced by a thin, very stably stratified interface. 

The convection on both sides continues to gradually erode the remaining interface which disappears around $t \approx 17000$. This erosion mechanism will be explored in more detail in a future publication. In the final, statistically-stationary stage of the simulation, the domain is again fully-convective, and has a constant density except for thin boundary layers. \\

\subsection{Low $R_0^{-1}$}
\begin{figure*}
\centerline{\includegraphics[width=0.9\textwidth]{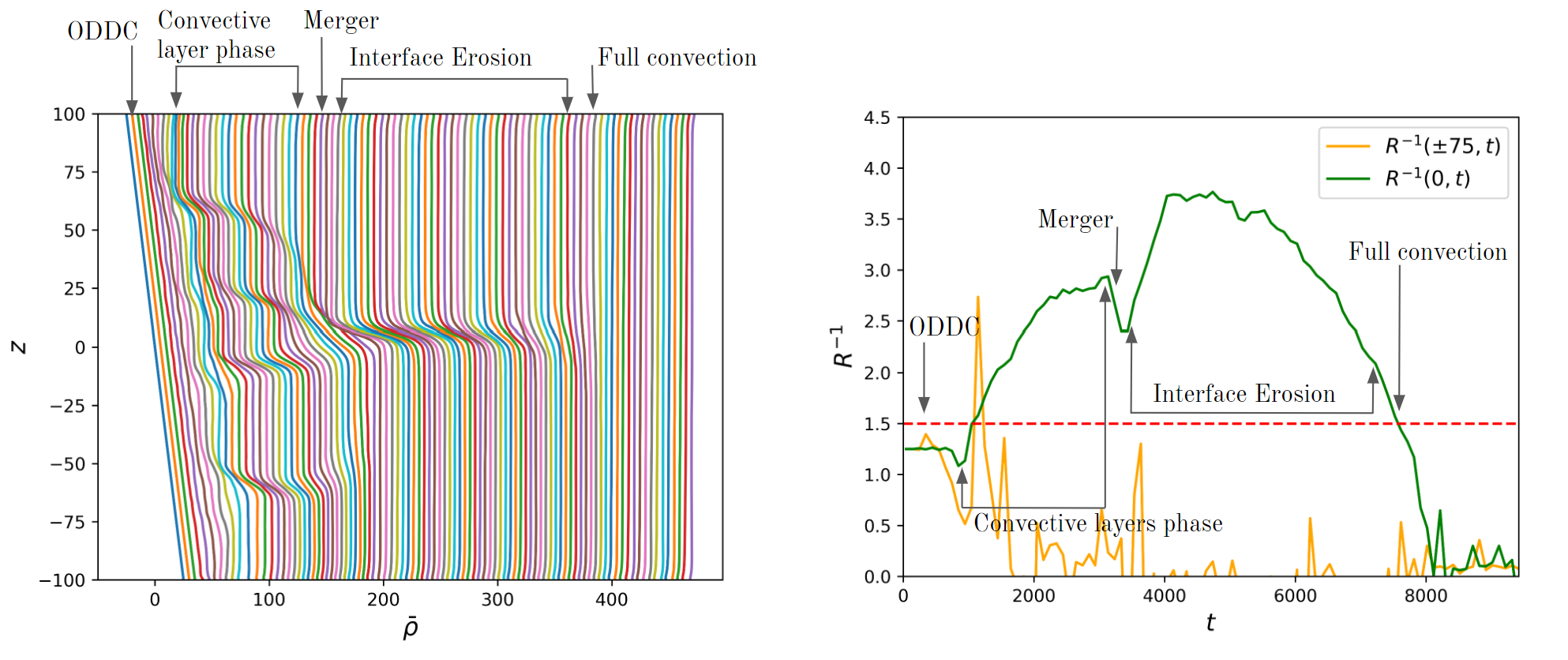}}
    \caption{As in Figure \ref{fig:case1}, for $Pr = \tau = 0.1$ and $R^{-1}_{0} = 1.25$. Left: Time-series of vertical density profiles. Each blue line corresponds to $\Delta t = 1,000$ time unit increments. The various evolutionary stages are marked with arrows. Right: Inverse density ratios vs time: $R^{-1}(0,t)$(green), and $R^{-1}(\pm75,t)$ (orange). The red horizontal dashed line shows $R_L^{-1} \simeq 1.5$, which is the layering threshold for $Pr = \tau = 0.1$.}
    \label{fig:case4}
    \end{figure*}

Finally, we present a simulation at very low $R^{-1}_{0}$ (but still $R^{-1}_{0}>1$). We see yet another completely different behavior in the evolution of the system. In Figure \ref{fig:case4}, we present a simulation with $Pr = \tau = 0.1$ and $R_{0}^{-1} = 1.25$. Similarly to other cases, the simulation starts with a linear density profile at $t = 0$ and evolves into a state of ODDC. This initial phase only lasts until $t \simeq  800$. After a short period of time, we  observe the formation of five convective layers in the domain at $t\approx 1000$. The convective layers are separated by thin strongly-stratified interfaces which are identified by sharp gradients in the density profiles. These layers are clearly formed by the $\gamma$-instability. Indeed, in this case, $R^{-1}(z,t)<R^{-1}_{L}$ everywhere at the start of the simulation, meaning that the $\gamma$-instability may drive the  formation of convective layers everywhere in the domain very early on.  Later, the number of convective layers is reduced from five to two, as a result of a  combination of interface erosion (e.g. lowest interface) and merger (e.g. top three interfaces). These processes happen quickly between $t \approx 1000$ and $ t \approx 3300$, resulting in a single interface in the middle of the domain separating two convection zones. Between $t\approx 3300$ and $t\approx 7500$, the interface is eroded over time by the two convective layers and eventually disappears at $t\approx7800$, leading once again to a fully convective chemically homogeneous domain.

To summarize, we have found that the long-term 
 evolution of this simple model system depends on  whether the $\gamma$-instability is triggered or not. 
In the first case, $R^{-1}(z,t)$ never decreases below $R^{-1}_{L}$, and the domain ultimately achieves a statistically-stationary state of ODDC. In all the other cases, the $\gamma$-instability drives the formation of convective layers as soon as $R^{-1}(z,t)$ drops below the threshold $R_{L}^{-1}$ somewhere in the domain. 
As in \citet{Woodal13}, these convective layers grow until they fill the entire domain, either by merging, or by entraining material from the neighboring ODDC regions. In other words, our numerical experiment has demonstrated that stably stratified ODDC regions can only survive for long periods of time provided $R^{-1}(z,t)$ never drops below $R_L^{-1}$.

\begin{table}
\begin{center}
\begin{tabular}{cccc}
   \hline
    $R_{0}^{-1 }$ & Resolution &  $R^{-1}(z,t)<R_{L}^{-1 }$ & Outcome\\
    \hline
     $Pr = \tau = 0.1$ & $(R_L^{-1} \simeq 1.5)$ & \\ 
         4.0 & 64$\times$64$\times$128 & no &  ODDC  \\ 
         3.0 & 64$\times$64$\times$128 & no &  ODDC  \\ 
         2.5 & 64$\times$64$\times$128 & yes, locally & CCL   \\ 
         2.0 & 64$\times$64$\times$128 & yes, locally &  BCL  \\ 
         1.75 & 128$\times$128$\times$256 & yes, locally  & BCL   \\ 
         1.4 & 128$\times$128$\times$256 & yes, locally  & BCL   \\ 
         1.25 & 128$\times$128$\times$256 & yes & $\gamma$-CL   \\ 
\hline 
     $Pr = \tau = 0.3$ & $(R_L^{-1} \simeq 1.25)$ & \\ 
         1.75 & 64$\times$64$\times$128 &  no & ODDC \\
         1.55 & 64$\times$64$\times$128 &  yes, locally & CCL \\
         1.4 & 128$\times$128$\times$256 & yes, locally &  BCL  \\ 
         1.3 & 128$\times$128$\times$256 & yes, locally & BCL   \\ 
         1.25 & 128$\times$128$\times$256 & yes, locally & BCL   \\ 
         1.15 & 128$\times$128$\times$256 & yes, locally & BCL   \\ 
         1.10 & 128$\times$128$\times$256 & yes & $\gamma$-CL   \\ 
\hline
\end{tabular}
\caption{List of simulations for $Pr = \tau = 0.1$ and $Pr = \tau = 0.3$, and their outcomes. The first column shows the initial inverse density ratio. The second column shows the adopted resolution in terms of numbers of modes, see section \ref{sec:setup}. The third column shows whether $R^{-1}(z,t)$ ever drops below the threshold, $R^{-1}_{L}$, at any point in the simulation. Finally, the fourth column indicates the outcome: "ODDC" is a statistically stationary ODDC state, `CCL' means that a layer forms in the center of the domain first, `BCL' means that layers form near the boundaries first, `$\gamma$-CL'indicates a simulation with a multitude of layers that form simultaneously. \label{tab:simulations}}
\end{center}
\end{table}

\section{Discussion \& Conclusions}
\label{sec:discussion}

Given the idealized nature of the DNS conducted in the previous section, one may question the generality of this conclusion. However, we believe it is quite robust. First, note that while we have presented results for $Pr= \tau=0.1$, very similar results are obtained at other input parameters, in as much as we have observed the same sequence of possible cases at $Pr = \tau = 0.3$, with the same conclusion regarding the role of $R_L^{-1}$ (see Table \ref{tab:simulations}). Second, \citet{Radko2007} has demonstrated that a density staircase (formed of convective layers separated by thin stable interfaces) necessarily coarsens with time either when the density flux increases with layer height or when it decreases with interface stratification. Both of these conditions are satisfied in ODDC staircases formed by the $\gamma$-instability \citep{Woodal13, Molletal2017}, suggesting that the coarsening process observed in figure \ref{fig:case4} (and more generally in any weakly-stratified region) is universal and will not stop until the entire region is fully convective. The robustness of cases illustrated in figures \ref{fig:case2} and \ref{fig:case3}, where a convective layer grows by eroding the nearby ODDC layer(s), is somewhat less clear but remains likely because the entrainment process takes place on dynamical timescales, while the restratification of the ODDC layer (necessary to preserve it) would take place on much longer evolutionary timescales. Finally, giant planets in our solar system are rapid rotators, and so one may wonder how rotation affects the $\gamma$-instability and the evolution of convective layers. \citet{MollGaraud2017} showed that the $\gamma$-instability is not {\it directly} affected by rotation, but the flux ratio $\gamma^{-1}$ itself, and its dependence on $R^{-1}$, may both depend on the planet's rotation rate. As such, while our general conclusion continues to hold in the presence of rotation, the critical inverse density ratio for layer formation $R_L^{-1}$ could be a function of the rotation rate for rapidly-rotating planets. A thorough study of the effects of rotation on the flux ratio $\gamma^{-1}$ is therefore needed. In addition, \citet{Fuentes_2023} showed that rotation can slow down the erosion rate of stably-stratified region adjacent to convective ones. This could slow down, but cannot completely halt, the growth of the fully convective zones once they have been created. 

 Of course, one should not {\it directly} compare the detailed evolution of the  model system studied in this paper to the evolution of a planet's diffuse core, because the boundary conditions we have used impose {\it fixed} and {\it equal} fluxes of heat and composition on both boundaries, while in reality this would certainly not be the case. In particular, depending on where we assume these boundaries are, the top and bottom fluxes are likely to differ from one another and evolve with time on the planetary evolution timescale \citep{hindman2023dwindling}. However, we have demonstrated that the development of the $\gamma$-instability is a local process, which is independent of what happens at the boundaries, and only depends on the local stratification of that core at any point in time.   
  
Because of this, the following conclusions hold:
\begin{enumerate}
    \item We confirm the result of \citet{Woodal13} and \citet{Mirouh2012} that the $\gamma$-instability plays a  fundamental role in the dynamics of stably-stratified regions with a super-adiabatic temperature gradient. More specifically, we have found that the $\gamma$-instability triggers convective layer formation locally as soon as the local inverse density ratio, $R^{-1}(z,t)$ (see eq. \ref{eq:localR}), drops below $R^{-1}_{L}(Pr,\tau)$ (see eq. \ref{eq:r_l}). Only
    regions with $R^{-1}(z,t)> R_{L}^{-1}, \forall z,t$ can remain in a long-term state of ODDC. This, incidentally, shows the importance of having good estimates for both $Pr$ and $\tau$ \citep[cf.][]{french2019kronoseismology,Preising} as well as as a good model for $R_{L}^{-1}(Pr, \tau)$ \citep{Mirouh2012}. If the planet is rapidly rotating,  $R_{L}^{-1}$ may also depend on the rotation rate.
    \item Our results, combined with those of \citet{Radko2007}, challenge the standard paradigm of compositionally stratified regions in giant planets which pictures them as a stack of long-lived, well-mixed, shallow convective layers, separated by thin diffusive interfaces \citep{Stevenson1982}. This has not been seen as an outcome of our simulations. At the moderate stratifications where they spontaneously form, convective layers always seem to grow and merge, until the convection zone occupies the entire domain. These processes happen rapidly and irreversibly. We believe that rapid rotation might slow down, but cannot completely hinder the process.    
  \item The fact that Saturn is observed to have an extended stably-stratified core suggests that $R^{-1}(z,t)$ has never decreased below the threshold $R_{L}^{-1}$ in that core  throughout the evolution of the planet. This might help constrain formation and evolution models of Saturn.   
\end{enumerate}
There are a number of further investigations in planetary modeling that should be explored, given these results.  We have strong evidence that Saturn's dilute core today is stably stratified \citep{fuller2014saturn,MankovichFuller2021,fortney2023saturns}. Our work here suggests that this stable stratification must have already been established in Saturn during the planet's formation. This places constraints on the central temperature (and the corresponding temperature gradient in the dilute core) at that time, as well as at any time between formation until today, to ensure that the $\gamma$-instability is never triggered.  These temperature constraints directly inform static structure models today as well as thermal evolution models that aim to reproduce the planet's measured luminosity at its current age.

Without seismic constraints for Jupiter, we cannot as easily constrain the extent and thermal state of its dilute core.  One could imagine that like Saturn, Jupiter's dilute core is also stably stratified. This scenario is supported by the suggestion that internal gravity waves trapped in such a core produce enough of a gravitational signature to perturb the gravity field of tides raised by the satellite Io \citep{idini2022gravitational}. Alternatively, perhaps the higher initial thermal energy of the more massive Jupiter, and corresponding higher initial central temperature, have put the planet in a different regime, having triggered convection sometimes in the past so that the dilute core today is fully or partially homogenized. 

Further analysis is required to translate the conditions of instability presented here to the conditions expected in the interiors of Jupiter and Saturn. In the giant planets, such work must consider the nontrivial calculation of how material properties change with depth, including the adiabatic or super-adiabatic behavior of the mixture of H-He fluid with heavy elements and the uncertainties therein. Future research will aim to place new constraints on the range of core temperatures, strength of compositional gradients, and radial extent of dilute cores in Jupiter and Saturn, to assess what present and past interior structures are possible  given the mixing arguments developed here.




\begin{acknowledgments}
A.T. and P.G. gratefully acknowledge help from the 'Dedalus' development team and support from National Science Foundation grant AST-1908338. B.I. is supported by the University of California President's Postdoctoral Fellowship Program. Simulations were performed using the Extreme Science and Engineering Discovery Environment (XSEDE) Expanse supercomputer at the San Diego Supercomputer Center through allocation TG-PHY210050, as well as the Lux supercomputer at UC Santa Cruz, funded by NSF MRI grant AST-1828315.
\end{acknowledgments}



\end{document}